# Hot Carrier Extraction Using Energy Selective Contacts and Its Impact on the Limiting Efficiency of a Hot Carrier Solar Cell

Steven C. Limpert and Stephen P. Bremner
*University of New South Wales School of Photovoltaic and Renewable Energy Engineering, Sydney, NSW, 2033, Australia*

Extraction of charge carriers from a hot carrier solar cell using energy selective contacts, and the impact on limiting power conversion efficiency is analyzed. It is shown that assuming isentropic conversion of carrier heat into voltage implies zero power output at all operating points. Under conditions of power output, lower voltages than in the isentropic case are obtained due to the irreversible entropy increase associated with carrier flow. This lowers the limiting power conversion efficiency of a hot carrier solar cell.

The hot carrier solar cell (HCSC) is an advanced concept, or third generation, photovoltaic (PV) device offering a potential pathway to efficiencies in excess of the conventional homo-junction limit[1]. The basis of operation of a HCSC is to utilize the temperature gradient between the lattice and photo-generated carriers to obtain larger voltages than are obtainable in a uniform temperature device[2]. A model device consists of an absorber and energy selective contacts (ESCs) that allow carriers to be extracted from the absorber at specific energies or over certain energy ranges as shown in FIG. 1.

Analysis of energy losses in single band gap PV devices by Markvart[3] crucially shows that eliminating the carrier cooling mechanism is thermodynamically consistent. However, a method for making use of the temperature gradient between the lattice and the photo-generated carriers has been a topic of debate. Previous theoretical studies of HCSCs have focused on the behavior of carriers in the absorber and the nature of the absorber chemical potentials[2,4,5]. A model is typically used for the ESCs in which carrier extraction occurs isentropically and carrier energy is converted into voltage at Carnot efficiency.

The purpose of this letter is to highlight that the assumption of isentropic carrier extraction and conversion of energy into voltage at the Carnot efficiency leads to an inconsistency in the model of a HCSC and that this assumption will lead to an overestimation in limiting power conversion efficiency. It is shown that isentropic carrier extraction corresponds to the open circuit case of the HCSC and that for all other voltage operating points, the chemical potential difference of extracted carriers must be less than that in the isentropic case.

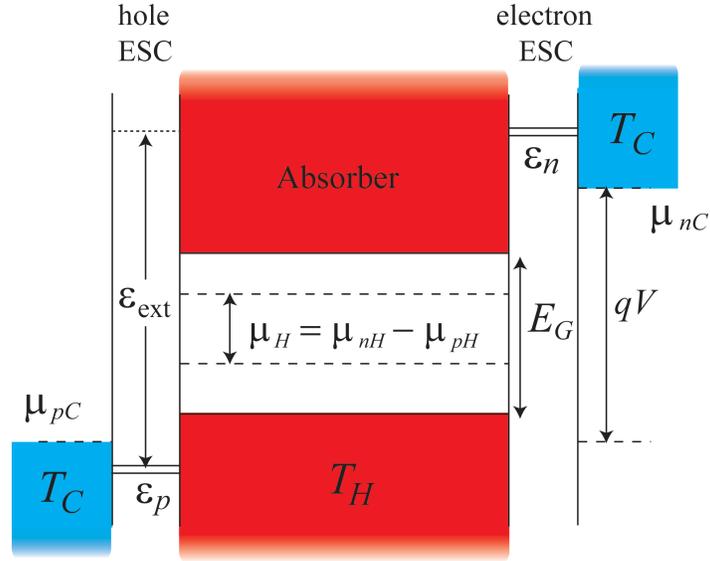

**FIG. 1:** General scheme for a hot carrier solar cell showing energy selective contacts that allow extraction of carriers from an absorber with carrier temperature $T_H$ to contacts with carrier temperature $T_C$.

The assumption of isentropic carrier extraction and conversion of heat into voltage in the first published work on hot carrier solar cells[2] resulted in the expression for the voltage provided by a hot carrier solar cell being given as

$$qV \equiv \mu_{nC} - \mu_{pC} = \varepsilon_{ext}\left(1 - \frac{T_C}{T_H}\right) + \mu_H \frac{T_C}{T_H} \qquad (1)$$

where $T_H$ and $T_C$ are the temperatures of the carriers in the absorber and the ESC respectively, and $\varepsilon_{ext} \equiv \varepsilon_n - \varepsilon_p$ where $\varepsilon_n$ is the energy of the electron ESC and $\varepsilon_p$ is the energy of the hole ESC.

By analyzing the entropy associated with the extraction of one photo-generated electron-hole pair from the absorber, the origin of this expression and the assumptions that underpin it can be illuminated. To begin, the change in entropy of an electron when it moves from the absorber, through the ESC to the contact region is given by the difference in entropy of the electron in its final and initial states. If we take the hot absorber and the cold contact to be thermally isolated electron reservoirs, the entropy associated with an electron being transferred from absorber to contact can be written as[6]:

$$\Delta S_n = \frac{\varepsilon_n - \mu_{nC}}{T_C} - \frac{\varepsilon_n - \mu_{nH}}{T_H} \quad (2)$$

where $\mu_{nC}$ and $\mu_{nH}$ are the chemical potentials of the electrons in the cold contact and the hot absorber, respectively. It is assumed that the transfer of the electron does not change $\mu_{nH}$ or $\mu_{nC}$, that the electron energy does not change during the transfer and that there are no equilibration losses when the electron reaches the contact region. For the hole extraction process, similar reasoning leads to the following condition:

$$\Delta S_p = \frac{\mu_{pC} - \varepsilon_p}{T_C} - \frac{\mu_{pH} - \varepsilon_p}{T_H} \quad (3)$$

where $\mu_{pC}$ and $\mu_{pH}$ are the chemical potentials of the holes in the cold contact and the hot absorber respectively.

In order for the overall electron-hole pair extraction to be isentropic, both the electron extraction and hole extraction processes must be individually isentropic. It is therefore helpful to analyze the ESCs in isolation. Concentrating first on the case of electrons and setting equation (2) to zero, an expression for the chemical potential difference that ensures isentropic transport of electrons between absorber and the contact is found:

$$\Delta \mu_n \equiv \mu_{nC} - \mu_{nH} = \left(1 - \frac{T_C}{T_H}\right)(\varepsilon_n - \mu_{nH}). \quad (4)$$

We can re-cast (4) to express the extraction energy at which extraction is isentropic:

$$\varepsilon_n = \frac{T_H \mu_{nC} - T_C \mu_{nH}}{T_H - T_C} \quad (5)$$

This is the exact same condition found by Humphrey et al.[6] for a quantum thermoelectric device operating at the Carnot limit between electron reservoirs at temperatures $T_H$ and $T_C$. As highlighted in that work, operation at this point results in zero net carrier transfer between hot and cold reservoirs. In other words, there is no current and therefore no power output under these conditions.

This result is not surprising in light of an important insight: the HCSC is a type of particle exchange heat engine[7], where heat transfer occurs by particle exchange, and work is extracted from the heat transfer facilitated by this particle exchange. Thermoelectric devices deliver work using the same basic operating principle. The basis of this particle exchange process at the electron ESC is indicated in FIG. 2, where the long tail of occupied states for the hot carrier population results in a net flow of 'hot' electrons from the absorber to the contact at low $\Delta\mu_n$ (this is indicated by the arrow in the figure to the left). As $\Delta\mu_n$ increases, the occupation probabilities get closer in value until at a particular value of $\Delta\mu_n$, they are exactly equal. When this $\Delta\mu_n$ value is reached, neither flow direction is favored (this is indicated by the double headed arrow) and the net carrier transfer is zero. It is at this point, and this point only, that isentropic transport is possible. Increasing $\Delta\mu_n$ further than this point results in net flow of carriers from the cold contact region into the hot absorber. This leads to the following statement: for a given absorber temperature, $T_H$, contact temperature, $T_C$, and extraction energy, $\varepsilon_n$, isentropic transport of particles occurs only at the $\Delta\mu_n$ value for which the occupation of carrier states in the absorber and contact regions are exactly equal.[6]

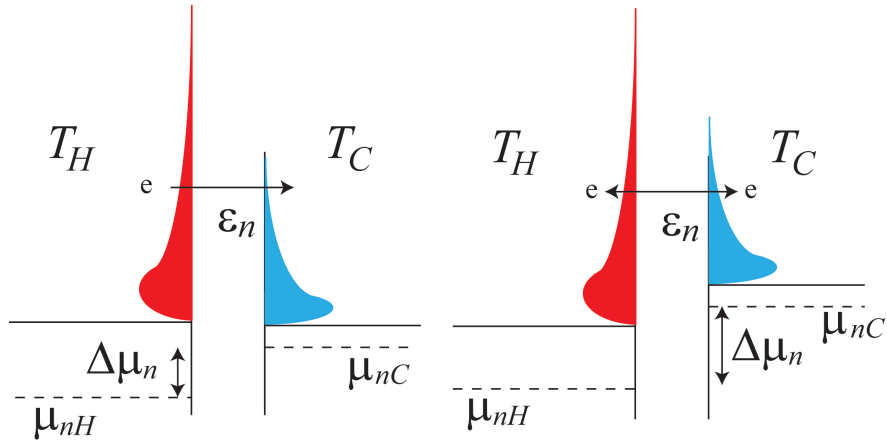

**FIG. 2:** Electron Energy Selective Contact showing that the difference in occupation is the driving force behind the current output of the HCSC. When the occupations are equal, there is no net current flow.

Similar considerations for the case of the hole ESC leads to a corresponding chemical potential term for isentropic transport of holes from absorber to the contact:

$$\Delta\mu_p \equiv \mu_{pH} - \mu_{pC} = \left(1 - \frac{T_C}{T_H}\right)\left(\mu_{pH} - \varepsilon_p\right). \quad (6)$$

Again, recasting this expression to find the extraction energy for isentropic transport, we have:

$$\varepsilon_p = \frac{T_H \mu_{pC} - T_C \mu_{pH}}{T_H - T_C}. \quad (7)$$

It is straightforward to prove that a similar equality between the absorber and contact hole occupation functions occurs at this point. Adding together (4) and (6), and making use of the definitions $\mu_H \equiv \mu_{nH} - \mu_{pH}$ and $\varepsilon_{ext} \equiv \varepsilon_n - \varepsilon_p$, we arrive back at (1), an expression which is now seen to be best interpreted as describing the open circuit voltage for a HCSC with an absorber temperature, $T_H$, contact temperature, $T_C$, extraction energy difference, $\varepsilon_{ext}$, and hot carrier chemical potential splitting, $\mu_H$.

Having established that the isentropic extraction voltage always corresponds to the open circuit voltage for a given $T_H$, $T_C$, $\varepsilon_{ext}$ and $\mu_H$, in order to derive a voltage expression which is meaningful at other points on the current-voltage (IV) curve, we must return to the more general case of the entropy increase associated with electron and hole extraction expressed in equations (2) and (3). If we denote the sum of these entropy terms as $\Delta S_{ESC}$, the following expression is found that relates the voltage produced by the hot carrier solar cell to the energy levels of extraction, the carrier temperatures, the splitting of the hot carrier chemical potentials and the increase in the entropy of the carriers that occurs during the extraction process:

$$qV \equiv \mu_{nC} - \mu_{pC} = \varepsilon_{ext}\left(1 - \frac{T_C}{T_H}\right) + \mu_H \frac{T_C}{T_H} - T_C \Delta S_{ESC}. \quad (8)$$

This expression holds regardless of the value or sign of $\mu_H$, making questions over the exact nature of this chemical potential difference a moot point. Previous work by Würfel[4,5] suggested the term $\mu_H$ is forced to be zero in order to make a physically sensible HCSC model where Auger processes between the hot carriers are included. Such considerations are the beyond the scope of this paper.

It is important to stress that $\Delta S_{ESC}$ is not zero by default, and that it corresponds to the irreversible entropy generation associated with current flow through the HCSC. In all scenarios in which there is current flow (i.e. a process with time directionality), $\Delta S_{ESC}$ must be positive as entropy must increase in time. Another way to think about this is to consider that carriers move in order to increase their entropy, carriers cannot move in such a way as to decrease their entropy, and when carriers have no opportunity to increase their entropy, they do not move at all. That is to say, when the entropy of carriers is not increasing, those carriers are not moving. Such a scenario occurs when $\Delta S_{ESC}$ is zero: no current flows and the device is operating at its open-circuit voltage point. At this point, the device is operating at the best possible energy conversion efficiency as the produced voltage is a maximum, but it is operating at zero power conversion efficiency, as there is no power produced. Similar scenarios in heat engines in which the maximum efficiency differs from the efficiency at maximum power have been discussed in the literature[6,8].

One further observation is that the presence of $\Delta S_{ESC}$ enables a case to be described in which the absorber temperature is greater than the contact temperature at short-circuit. This was not possible in some previous HCSC models[4,5]. A difference in temperature between the absorber and the contacts at short-circuit makes intuitive physical sense as it is the temperature difference which provides the driving force for carrier flow between the absorber and contact via the ESC. An implication of a temperature difference at short-circuit is that the short circuit current possible from the HCSC will be reduced as there will be emission from the absorber above the ambient level resulting in less carriers available to be extracted as current. The full implications of this updated expression for the voltage of a HCSC using ESCs need to be explored further as it will also mean the design of the extraction energy for the HCSC must no longer be based on the goal isentropic extraction.

In conclusion, analysis of the process of extracting charge carriers from a hot carrier solar cell using an energy selective contact scheme has shown that assuming isentropic carrier extraction implies zero power output at all operating points of device. Under conditions of power output, lower voltages than in the isentropic case are obtained due to the irreversible entropy increase associated with carrier flow. This lowers the limiting power conversion efficiency of a hot carrier solar cell. Further analysis of the carrier extraction process in ESCs as part of a HCSC is needed to better understand the conversion of carrier heat into voltage. The design and optimization of power conversion devices that operate based upon temperature gradients has been extensively investigated in the field of quantum thermoelectrics,[9-17] which may provide insight into how to design ESCs for hot carrier solar cells.

This work has been partially supported by the Australian Government through the Australian Renewable Energy Agency (ARENA). Responsibility for the views, information, or advice expressed herein is not accepted by the Australian Government. Steven Limpert acknowledges the financial support of the Australian-American Fulbright Commission.